\documentclass{article}
\usepackage{graphicx} 
\usepackage{hyperref}
\usepackage[numbers]{natbib}
\usepackage[scale=0.71]{geometry}

\title{Arctic Inference with Shift Parallelism: Fast and Efficient Open Source Inference System for Enterprise AI}
\author{Samyam Rajbhandari, Mert Hidayetoglu, Aurick Qiao, Ye Wang,\\ Juncheng Yang, Jeff Rasley, Michael Wyatt, and Yuxiong He
\\
\\
Snowflake AI Research}
\date{}

\begin{document}

\maketitle

\begin{figure}[h]
    \centering
    \includegraphics[width=\linewidth]{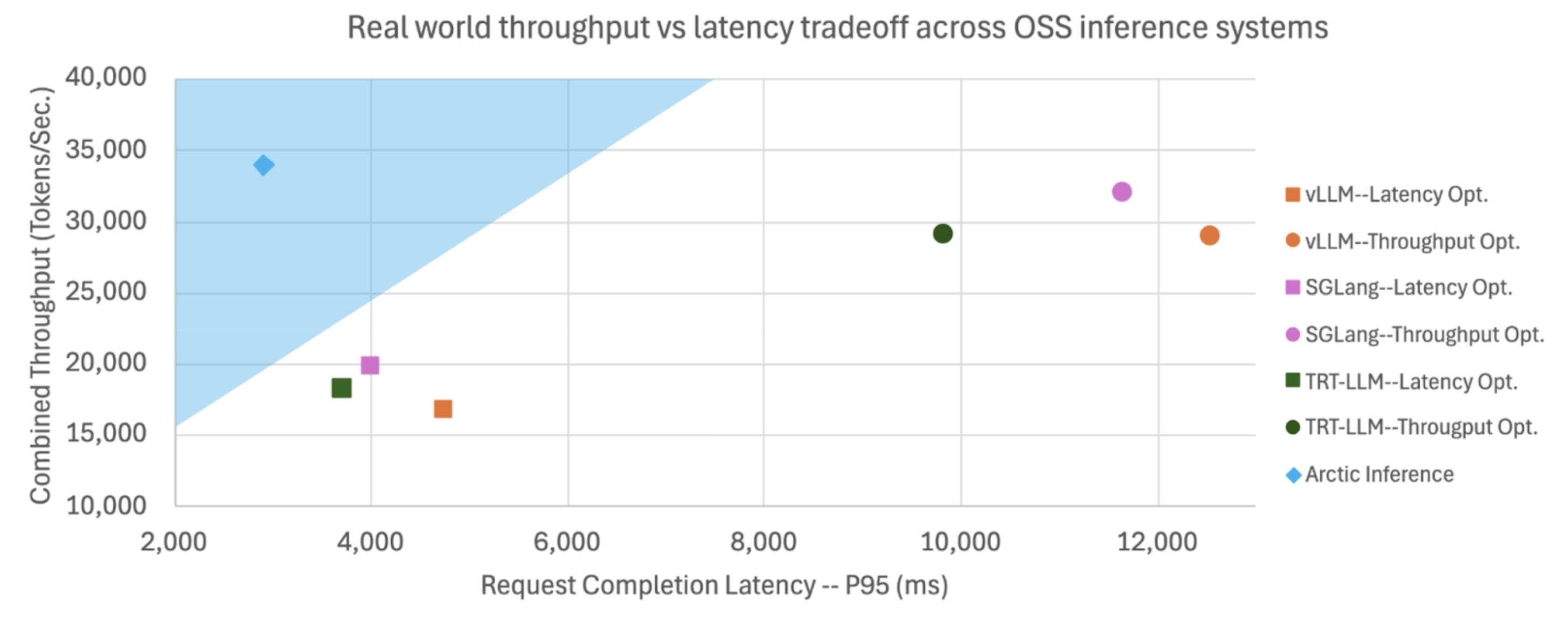}
    \caption{Arctic Inference achieves highest throughput and lowest latency for Llama 3.3 70B across open source inference frameworks.}
    \label{fig:hero2}
\end{figure}

\subsection*{Abstract}
Inference is now the dominant AI workload, yet existing systems force trade-offs between latency, throughput, and cost. Arctic Inference, an open-source vLLM plugin from Snowflake AI Research, introduces Shift Parallelism, a dynamic parallelism strategy that adapts to real-world traffic while integrating speculative decoding, SwiftKV compute reduction, and optimized embedding inference. It achieves up to 3.4× faster request completion, 1.75× faster generation, and 1.6M tokens/sec per GPU for embeddings, outperforming both latency- and throughput-optimized deployments. Already powering Snowflake Cortex AI, Arctic Inference delivers state-of-the-art, cost-effective inference for enterprise AI and is now available to the community.

\section{Introduction}

Inference is becoming the dominant workload in AI, but today’s systems force developers to make costly trade-offs between low latency, high throughput and affordable deployment. 

\href{http://github.com/snowflakedb/ArcticInference}{Arctic Inference} \citep{arcticInferenceRepo} changes that. Built by Snowflake AI Research, it’s an open source vLLM~\cite{vllmrepo} plugin that brings Snowflake’s inference innovations to the community, delivering the fastest, most cost-effective open source inference for enterprise AI (see Figure~\ref{fig:hero2}). 

At the core of Arctic Inference is \textbf{Shift Parallelism}, a new parallelism strategy designed to dynamically adapt to real-world challenges and unpredictable traffic, simultaneously achieving maximum speed (lowest time to first token and time per output token) and high cost efficiency (high throughput) in a single deployment. 

In this paper~\footnote{The related blog can be found here: \href{https://www.snowflake.com/en/engineering-blog/arctic-inference-shift-parallelism/}{Arctic Inference Blog} \cite{shiftParallelBlog2025}.}, we’ll dive into Shift Parallelism and how the full suite of innovations in Arctic Inference (cutting-edge speculative decoding~\cite{specDecodeBlog2025}, compute reduction with SwiftKV~\cite{swiftkvblog} and optimized embedding inference \cite{embeddingBlog2025}) advance the state of the art for real-world enterprise AI.

\subsection*{Real-world results, one deployment}

For real-world generative AI workloads, Arctic Inference+vLLM in a single deployment, achieves:

\begin{itemize}
    \itemsep 0em 
    \item 3.4$\times$ faster request completion and 1.06$\times$ higher throughput compared to state-of-the-art (SoTA) throughput-optimized deployment
    \item 1.7$\times$ higher throughput and 1.28$\times$ faster request completion compared to SoTA latency-optimized deployment
    \item the elusive trifecta: 2.25$\times$ lower response time, 1.75$\times$ faster generation and on-par throughput compared to bespoke deployments optimized for each metric
\end{itemize}

For non-generative AI workloads, such as embeddings, Arctic Inference+vLLM delivers a whopping 1.6M toks/sec per GPU:

\begin{itemize}
    \itemsep 0em 
    \item 16x faster than vLLM on short sequences and 4.2$\times$ faster on long sequences
    \item 2.4$\times$ faster than Text Embeddings Inference (TEI \cite{teirepo}) on short sequences and at parity for long sequences
\end{itemize}

The performance claims are supported with detailed evaluation results later in this paper. More importantly, they’re already delivering real-world impact in production with Arctic Inference powers key workloads in Snowflake Cortex AI\footnote{Snowflake Llama models and embedding models in Snowflake Cortex AI.}.

Today we’re excited to open source Arctic Inference and Shift Parallelism for the broader AI community. (If you'd like to cite this work, please use the BibTeX reference at the bottom of this post.)

\section{Why today’s inference systems fall short}
Inference workloads are not like training. While training workloads are homogeneous across batches, real-world inference traffic is highly dynamic, experiencing bursty, unpredictable patterns. Furthermore, while training is throughput driven, real-world inference workloads care for three distinct metrics:

\begin{itemize}
    \itemsep 0em 
    \item \textbf{TTFT} (time to first token): fast initial response
    \item \textbf{TPOT} (time per output token): fast full generation
    \item \textbf{Throughput}: cost-efficient token serving at scale
\end{itemize}

Existing parallelism strategies such as tensor parallelism and data parallelism were originally designed for the homogeneous, batch-optimized world of training. In real-world inference, this means significant trade-offs are introduced.

\begin{figure}[h]
    \centering
    \includegraphics[width=\linewidth]{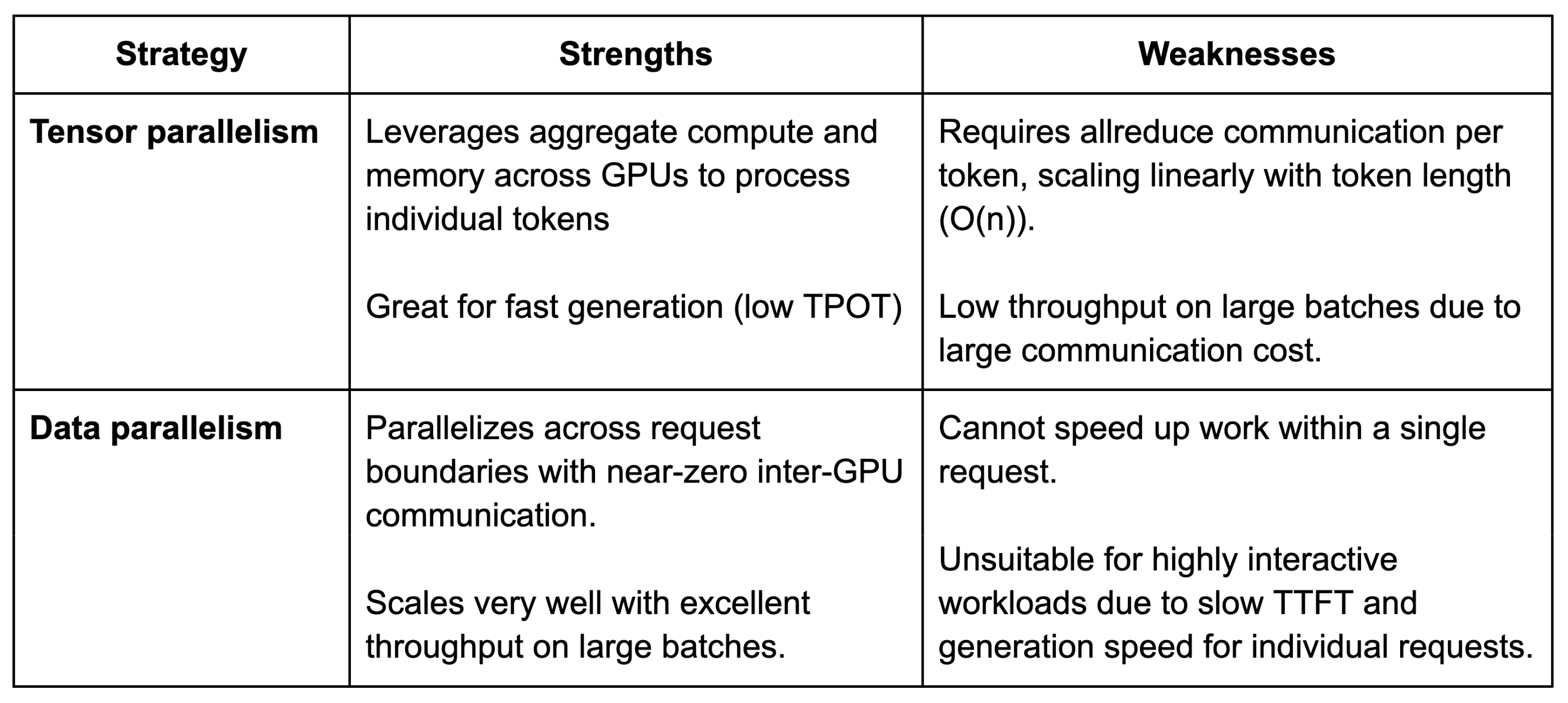}
\end{figure}

\subsection*{Why not combine them?}

Switching between tensor parallelism and data parallelism may seem obvious, but in practice, it's not viable. Their KV cache memory layouts are incompatible, and switching requires expensive data movement. Most teams resort to duplicating deployments: one for latency, one for throughput. This adds cost and complexity.

To overcome these KV cache limitations, Arctic Inference introduces a new strategy: \textbf{Arctic Sequence Parallelism} \cite{ulyssesBlog2025} (referenced as Arctic Ulysses in charts below).

Arctic Sequence Parallelism splits the input sequence across GPUs to parallelize work within a single request. Unlike tensor parallelism, it avoids costly token-wise communication ($O(n)$), while still achieving high GPU utilization. And because it shares a KV cache layout with Tensor parallelism, it’s the ideal counterpart for large-batch scenarios. See our \href{https://www.snowflake.com/en/engineering-blog/ulysses-low-latency-llm-inference/}{blog post} to learn more. 

With Arctic Sequence Parallelism in place, this means we can finally unify the best of both worlds.

\begin{figure}[t]
    \centering
    \includegraphics[width=\linewidth]{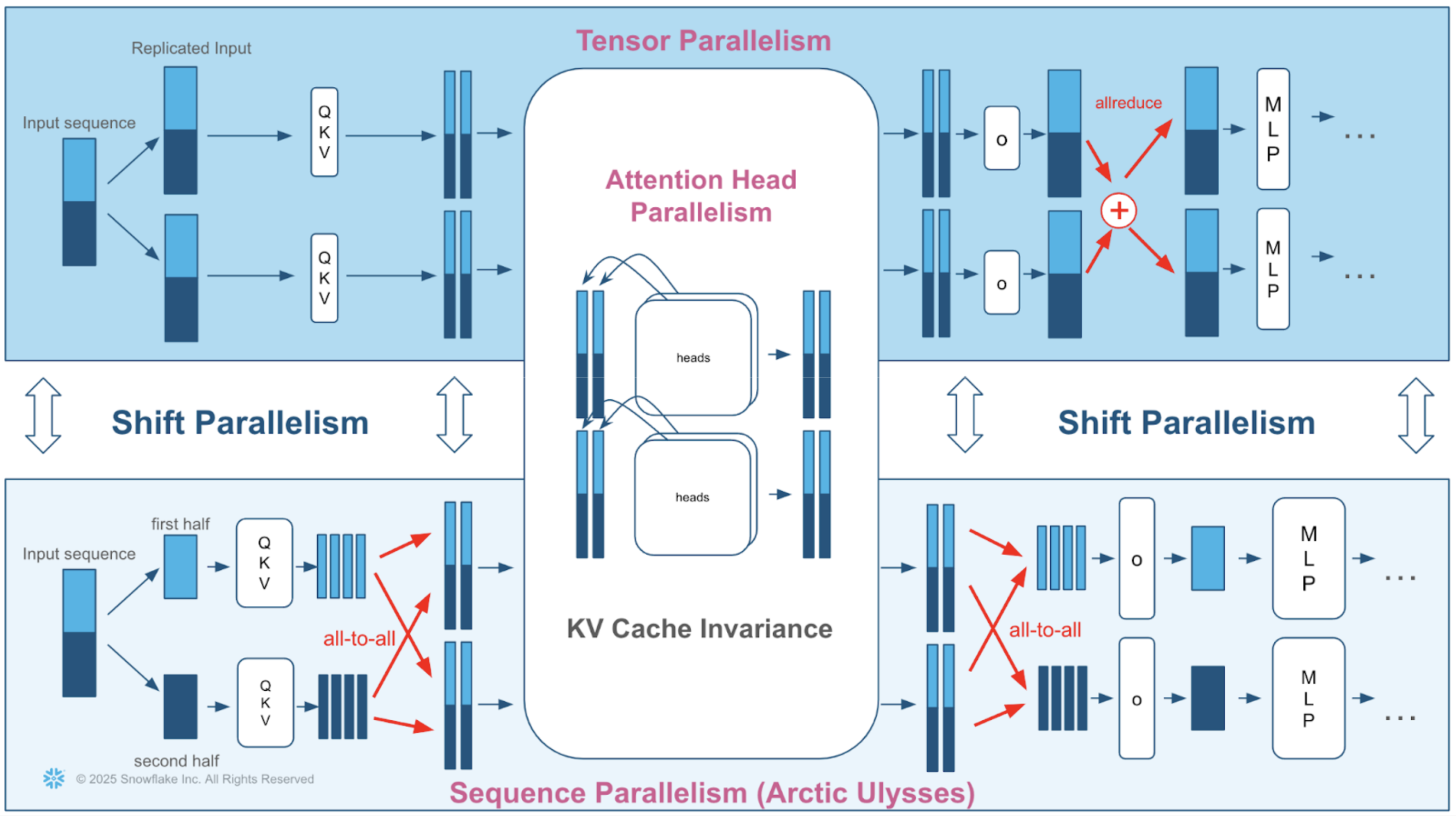}
    \caption{Shift Parallelism shifts between two modes: tensor parallelism and sequence parallelism (Arctic Ulysses).}
    \label{fig:shift_parallelism}
\end{figure}

\section{Introducing Shift Parallelism: One deployment, without the trade-offs}

Unlike traditional parallelism approaches that statically optimize for one of the three inference metrics---\textit{response latency, generation speed or cost efficiency}---\textbf{Shift Parallelism} dynamically adapts based on real-world traffic, delivering all three without requiring multiple deployments tuned for each. 

Shift Parallelism works by shifting between two best-in-class modes (see Figure~\ref{fig:shift_parallelism}):
\begin{itemize}
    \itemsep 0em 
    \item \textbf{Tensor parallelism} (TP) for small batches---maximizing output token generation speed (lower TPOT)
    \item \href{https://github.com/snowflakedb/ArcticInference/tree/main/projects/ulysses}{Arctic Sequence Parallelism} (SP) for large batches---minimizing TTFT and achieving near-optimal throughput
\end{itemize}

This is possible because the KV cache memory layout remains invariant between TP and SP, allowing Shift Parallelism to switch modes seamlessly, based on batch size and traffic patterns. More specifically, the KV cache layout does not change when changing SP and TP, as long as $\textrm{SP}\times\texttt{TP}$ equals \textrm{P}. 

This is shown concretely in Figure~\ref{fig:shift_parallelism}, where Shift Parallelism can switch between TP$=$2 and SP$=$2 (Arctic Ulysses) seamlessly across forward passes due to the KV Cache Invariance. The computation shown above is a single transformer layer with four attention heads running on two GPUs. In both TP$=$2 and SP$=$2, each GPU is computing two out of four attention heads. The computation and data mapping for the attention is identical across both TP and SP, allowing Shift Parallelism to switch between the two based on the size of the input.
 
Furthermore, by carefully mapping tensor parallel ranks to GPUs, we can ensure that the small parameter shards required on a GPU when using a large TP (TP$=$8, for example) are already part of the larger parameter shards present in that GPU needed to support a large SP (SP$=$8, for example).

\begin{figure}
    \centering
    \includegraphics[width=\linewidth]{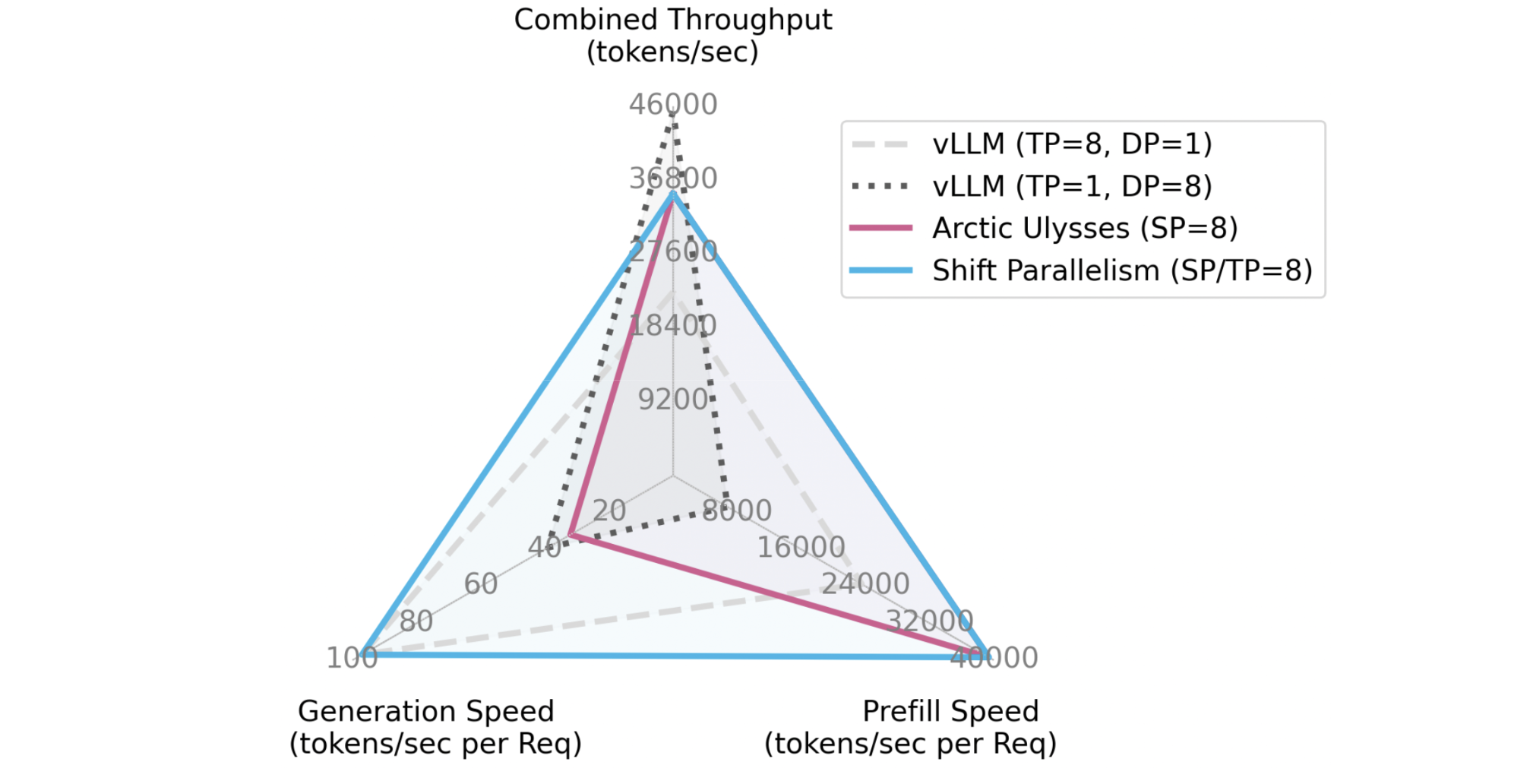}
    \caption{Latency vs. throughput trade-off between forms of parallelism for Llama 3.3 70B on 8$\times$H200 GPUs.}
    \label{fig:hero1_shift_parallel}
\end{figure}

\paragraph{The result:} a single deployment that optimizes simultaneously for TTFT, TPOT and combined throughput---mitigating the costly trade-offs that limit traditional inference systems (see Figure~\ref{fig:hero1_shift_parallel}).

With Shift Parallelism, enterprises are no longer forced to choose between a latency-optimized or throughput-optimized deployment. Table~\ref{tab:table_1} summarizes the latency-vs.-throughput trade-offs of the existing parallelism strategies discussed above and how Shift Parallelism mitigates them.

\begin{table}[h!]
    \centering
    \begin{tabular}{c}
         \raisebox{-\totalheight}{\includegraphics[width=0.85\textwidth]{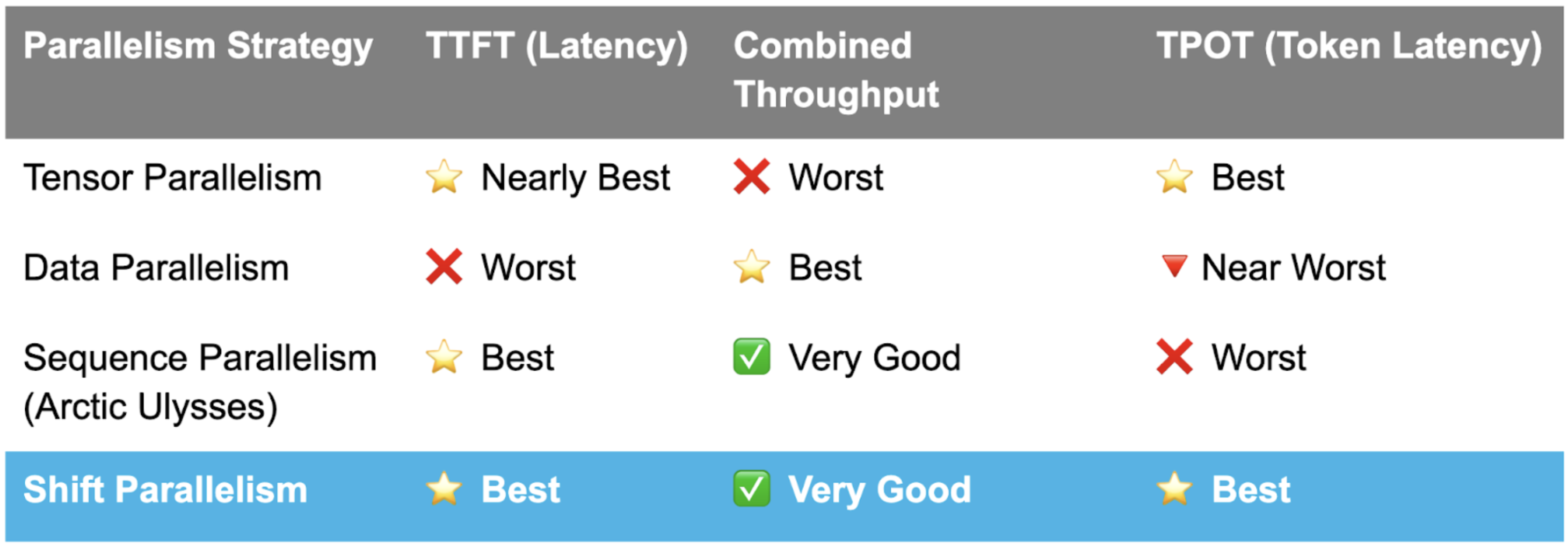}}
    \end{tabular}
    \caption{Latency vs. throughput trade-off between forms of parallelism (based on Figure~\ref{fig:hero1_shift_parallel}).}
    \label{tab:table_1}
\end{table}

\section{How Arctic Inference addresses real-world enterprise inference challenges}

Beyond Shift Parallelism, Arctic Inference includes a suite of advanced optimizations that target critical bottlenecks in enterprise AI workloads---from decoding and prefill inefficiencies to underoptimized embedding inference.

Below, we highlight how Arctic Inference solves key real-world challenges, with links to deeper technical blogs and papers.

\subsection*{Advancing speculative decoding for real-world generation}
Existing speculative decoding solutions are limited when it comes to real-world use: They do not leverage repetitive patterns in LLM generation; they lack optimized system implementations; and draft models such as EAGLE~\cite{eagle2025} in vLLM and sGLANG~\cite{sglangrepo} do not support inputs longer than 4,000 tokens, making them impractical.

Arctic Inference delivers the fastest speculative decoding in vLLM by combining suffix decoding and highly optimized light weight draft models, targeting repetitive and not repetitive generation patterns for real-world use cases. The result is up to \textit{4$\times$ faster generation for agentic workloads} (with repetitive patterns) and \textit{2.8$\times$ faster generation for conversational and coding workloads} (without repetitive patterns). \href{https://www.snowflake.com/en/engineering-blog/fast-speculative-decoding-vllm-arctic/}{Read more} on how this works.

\subsection*{Solving redundant prefill computation with SwiftKV}

In enterprise workloads, prefill often accounts for over 90\% of total compute. Yet open source frameworks such as vLLM~\cite{vllmpaper}, sGLANG~\cite{sglangpaper} and TRT-LLM~\cite{trtllmrepo} lack the optimizations to reduce this cost---leading to wasted resources on long inputs with minimal output.

SwiftKV \cite{swiftkvpaper} reuses hidden states from earlier transformer layers to eliminate redundant computation during KV cache generation---reducing prefill compute by up to 50\% without compromising accuracy. This results in up to \textit{2$\times$ higher throughput} for enterprise workloads with long prompts. To learn more about SwiftKV, please see our \href{https://arxiv.org/abs/2410.03960}{paper} and \href{https://www.snowflake.com/en/engineering-blog/swiftkv-llm-compute-reduction/}{blog post} on the topic

\subsection*{Solving embedding bottlenecks to enable over 1.5 million tokens/sec GPU performance}
Snowflake processes trillions of tokens per month across both real-time and batch embedding workloads. But when we benchmarked embedding models using vLLM, we uncovered performance bottlenecks---slow serialization, sequential tokenization and low GPU utilization---that left hardware severely underused. 

To fix this, we optimized Arctic Inference with vectorized serialization, parallel tokenization and multi-instance GPU execution. As a result, it delivers 16$\times$ faster embedding inference than vLLM on short sequences and 4.2$\times$ faster on long sequences, while outperforming TEI by 2.4$\times$ on short sequences and matching it on longer ones. Learn more in the blog post on \href{https://www.snowflake.com/en/engineering-blog/embedding-inference-arctic-16x-faster}{embedding throughput optimizations}.

\section{Bringing it all together: Proving Arctic Inference is best in class}

Here, we share core results demonstrating that Arctic Inference is the fastest and most cost-effective open source inference system for enterprise AI. (For technical readers, we also include a detailed evaluation methodology in the appendix at the end of this post.)

\subsection{Simultaneously, Arctic is currently the fastest and most cost-effective open-source inference system}

\begin{figure}
\includegraphics[width=\linewidth]{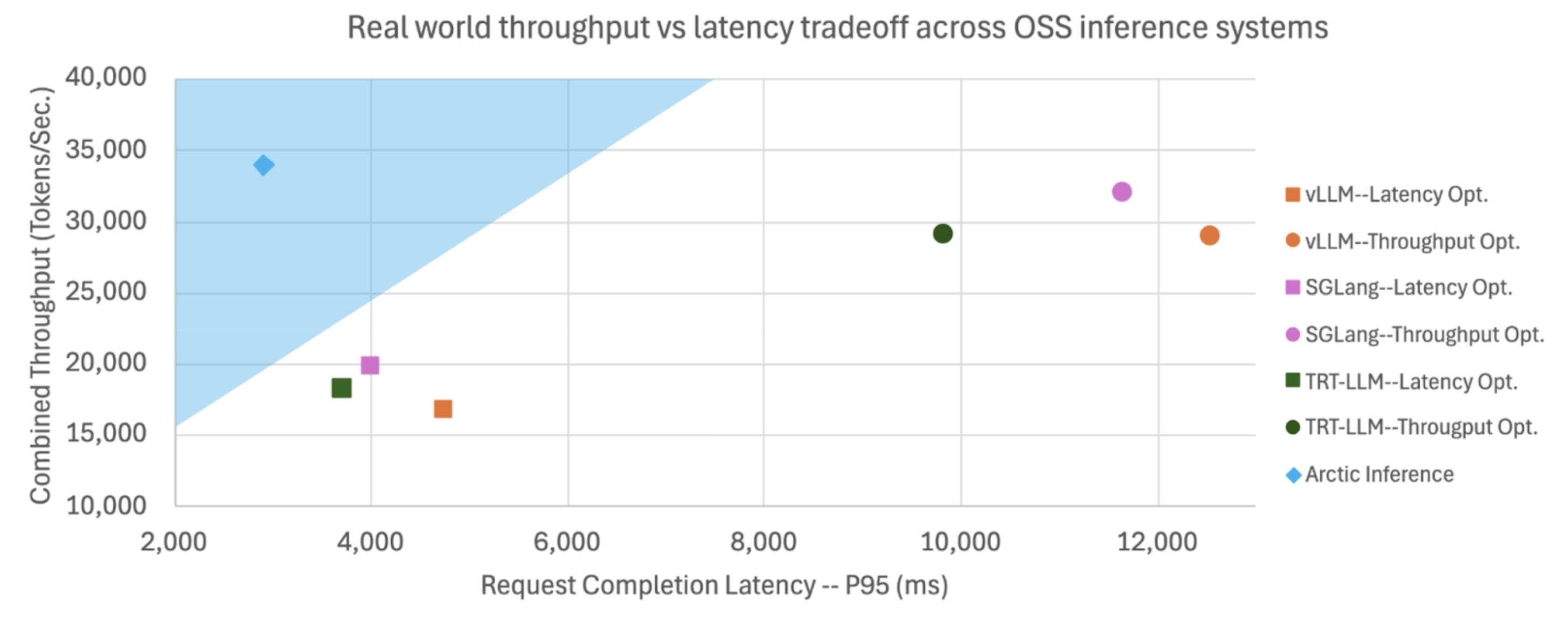}
\renewcommand\thefigure{\ref{fig:hero2}}
\caption{Arctic Inference achieves highest throughput and lowest latency for Llama 3.3 70B across open source inference frameworks.}
\end{figure}
\addtocounter{figure}{-1}

Figure~\ref{fig:hero2} shows that Arctic Inference simultaneously achieves highest throughput (lowest cost) and lowest completion time---all in one deployment---outperforming the best open source systems optimized for each metric individually\footnote{Latency-optimized and throughput-optimized configurations for vLLM, SGLang and TRT-LLM use TP$=$8 and DP$=$1 and TP$=$1 and DP$=$8, respectively, along with the best available speculative decoding for each framework. These experiments were run on data sets generated using real-world production traces to compute throughput, and a mixture of ShareGPT~\cite{sharegpt2023}, HumanEval~\cite{humaneval2021} and SWEBench~\cite{swebench2023} to measure latency. As a result, these results are representative of performance achievable in real-world deployments. For more details, see the evaluation methodology in the appendix.
}. More specifically, Arctic Inference combines Shift Parallelism with speculative decoding and SwiftKV to achieve:

\begin{itemize}
    \itemsep 0em 
    \item 3.4$\times$ faster request completion and 1.06$\times$ higher throughput compared to SoTA throughput-optimized deployment (TP$=$1, DP$=$8)
    \item 1.7$\times$ higher throughput and 1.28$\times$ faster request completion compared to SoTA latency-optimized deployment (TP$=$8, DP$=$1)
\end{itemize}

In Figure~\ref{fig:hero2}, latency-optimized and throughput-optimized configurations for vLLM, SGLang and TRT-LLM use $\textrm{TP}=8$ and $\textrm{DP}=1$ and $\textrm{TP}=1$ and $\textrm{DP}=8$, respectively, and the best speculative decoding solutions that were available for each of the frameworks (see the evaluation methodology in the appendix for details). These experiments were run on data sets generated using real-world production traces to compute throughput and a mixture of ShareGPT, HumanEval and SWEBench to measure latency. As a result, these results are representative of performance achievable in real-world deployments.

\subsection{Achieving the elusive trifecta: Quicker response, higher throughput and faster generation}

\begin{figure}
    \centering
    \includegraphics[width=0.8\linewidth]{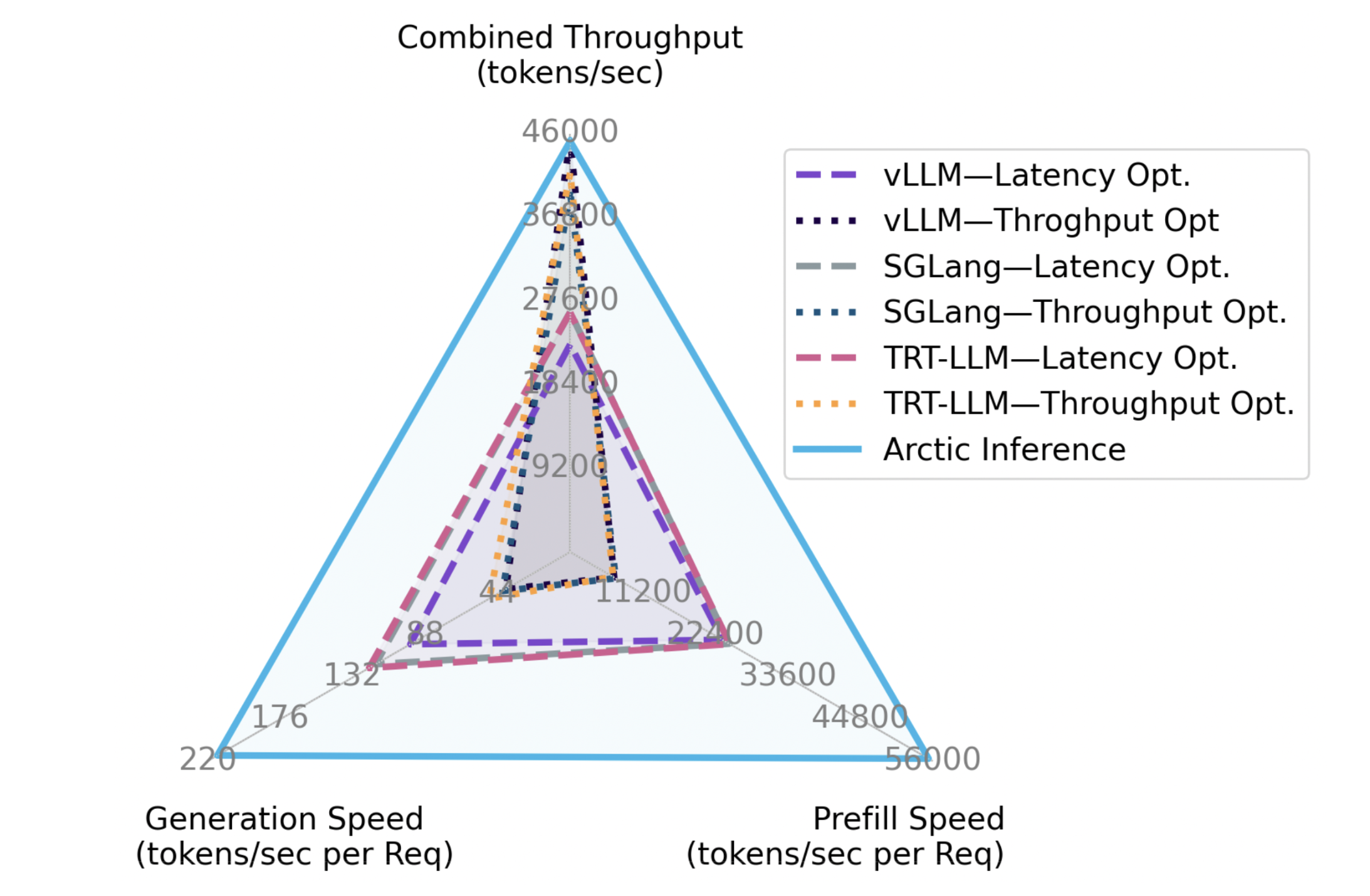}
    \caption{Arctic Inference simultaneously archives the fastest generation, prefill and combined throughput on Llama 3.3 70B running on 8$\times$H200.}
    \label{fig:hero1_arctic_inference}
\end{figure}

Response latency, generation speed and combined throughput are the three core pillars of inference system performance. Figure~\ref{fig:hero1_arctic_inference} shows that Arctic Inference outperforms the best open source systems optimized for each metric individually---achieving the elusive trifecta all in one deployment. More specifically, even when compared to the best deployment across vLLM, SGLang and TRT-LLM using bespoke configurations optimized for individual metrics, Arctic Inference with just a single deployment achieves: 

\begin{itemize}
    \itemsep 0em 
    \item 2.25$\times$ faster response time (prefill throughput per request) 
    \item 1.75$\times$ faster generation per request
    \item on-par combined throughput
\end{itemize}

This is possible due to the symbiosis between Shift Parallelism, optimized speculative decoding implementation and SwiftKV, which all work together in Arctic Inference. The combination of Shift Parallelism and speculative decoding enables Arctic Inference to achieve the fastest generation per request. Similarly, the combination of Shift Parallelism and SwiftKV enables Arctic Inference to achieve both the highest prefill speed, resulting in the fastest response times, and the highest throughput.

For details on the data sets used to produce these results, see the evaluation methodology in the appendix.

\subsection{16$\times$ faster throughput when scaling vLLM embeddings}

\begin{figure}
    \centering
    \includegraphics[width=\linewidth]{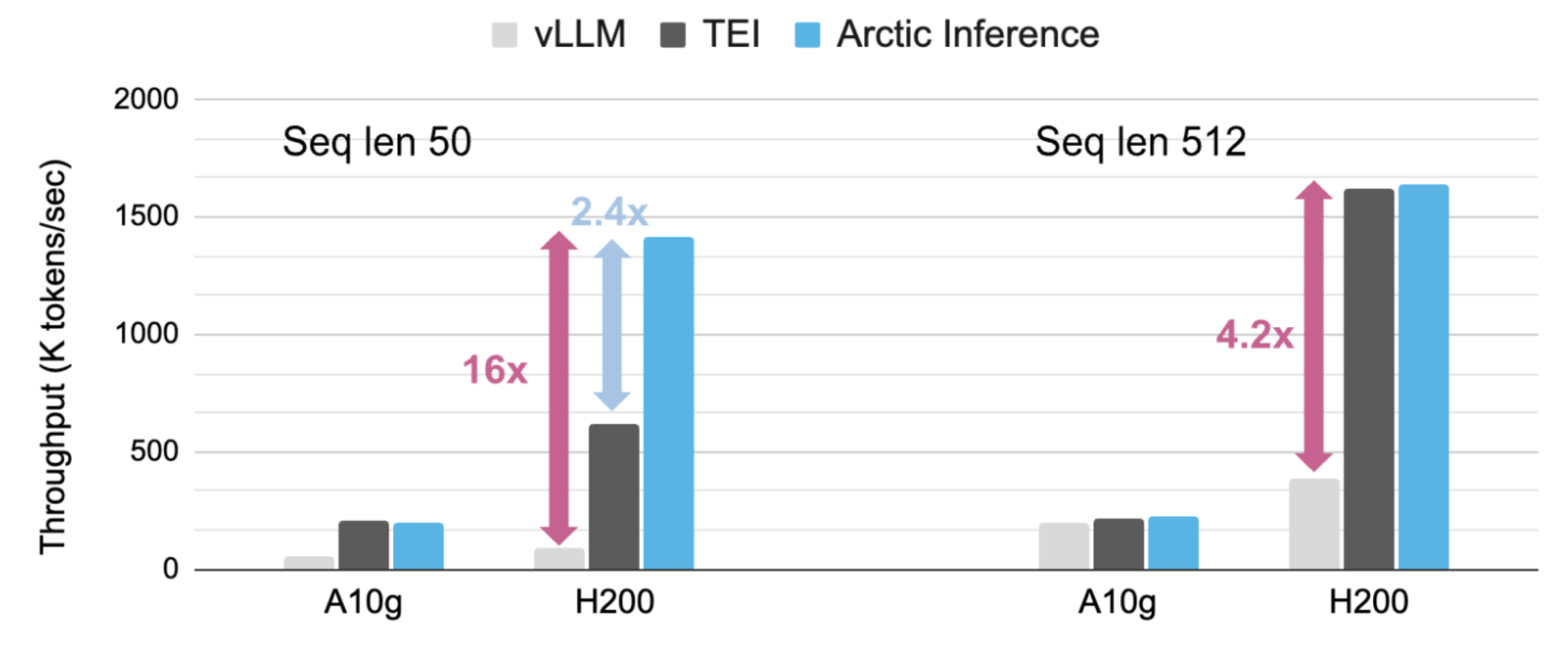}
    \caption{Arctic Inference establishes the new SoTA for embedding throughput performance, outperforming vLLM and TEI.}
    \label{fig:embedding}
\end{figure}

Figure~\ref{fig:embedding} shows that Arctic Inference can process over 1.4 million tokens per second not only for long sequences but also for short ones, which are notoriously difficult to optimize. 

By vectorizing data serialization and parallelizing tokenization, Arctic Inference helps ensure that the majority of computation time is spent on the actual embedding computation. 

As a result, Arctic Inference can achieve:
\begin{itemize}
    \itemsep 0em 
    \item 16$\times$ higher throughput than vLLM on short sequences and 4.2$\times$ higher throughput on long sequences
    \item 2.4$\times$ higher throughput than Text Embeddings Inference (TEI) on short sequences and on-parity for long sequences 
\end{itemize}

Furthermore, Arctic Inference supports running multiple instances of the same embedding model on a single GPU to allow better saturation of GPU resources when using small but powerful embedding models such as the \href{https://huggingface.co/collections/Snowflake/arctic-embed-661fd57d50fab5fc314e4c18}{snowflake-arctic-embed model family}. You can read more about this in our blog.

\subsection{Adapting to real-world traffic without a latency-throughput trade-off}

\begin{figure}[h]
    \centering
    \includegraphics[width=\linewidth]{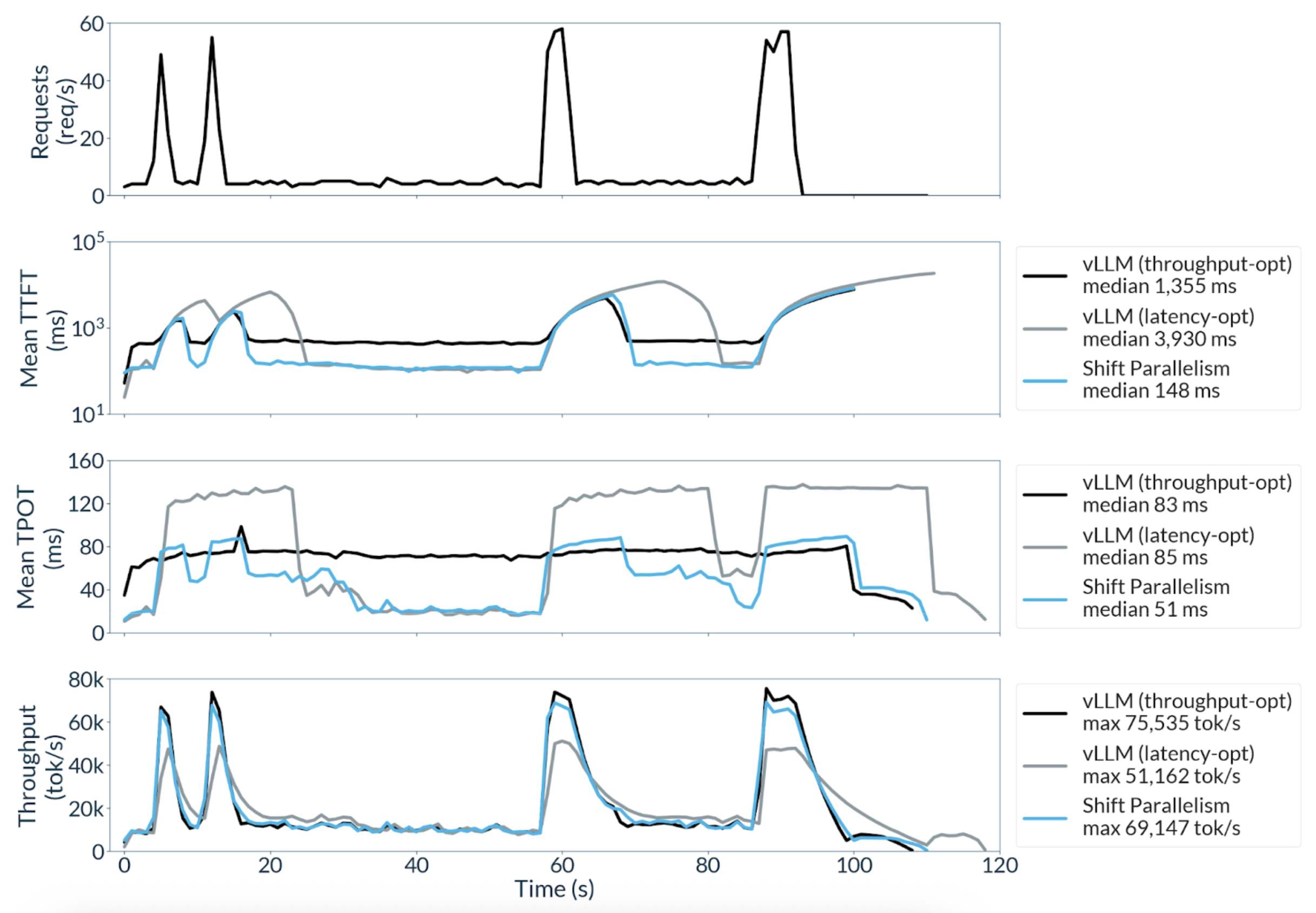}
    \caption{Shift Parallelism achieves the lowest response, fastest generation and near-optimal throughput under dynamic traffic.}
    \label{fig:dynamic}
\end{figure}

Figure~\ref{fig:dynamic} shows that Shift Parallelism can adapt to real-world traffic, simultaneously delivering the lowest response latency (TTFT), while achieving the fastest generation (lowest TPOT), and near-optimal cost efficiency (total throughput), compared to both throughput-optimized (DP only) and latency-optimized (TP only) solutions. More specifically, Shift Parallelism achieves:
\begin{itemize}
    \itemsep 0em 
    \item 9$\times$ reduction in median TTFT compared to the next best solution (1355ms → 148ms)
    \item 1.6$\times$ reduction in median TPOT compared to the next best solution (83ms → 51ms) 
    \item 1.6$\times$ reduction in median TPOT compared to the next best solution\footnote{vLLM does not allow measurements in real time, so the combined throughput as a function of time was obtained based on request start time, TTFT and generation throughput. As request arrival and first token response may not always align with the measurement time window, the calculated numbers are not always precise. However, since each parallelism config will have a similar margin of error, the relative measurements across different parallelism configurations are still very meaningful.} (83ms → 51ms) 
\end{itemize}

Max throughput regression less than 10\% compared to the best solution (75.5K → 69.1K toks/sec)
Here, Shift Parallelism dynamically shifts to using TP=8 when traffic is low, achieving the lowest TPOT, while switching to SP=8 when traffic increases, allowing for up to 1.35$\times$ higher throughput than $\textrm{TP}=8$ to avoid spikes in TTFT and TPOT.

\section*{Acknowledgments}
We would like to thank Jaeseong Lee and Gabriele Oliaro for their contributions to speculative decoding optimizations in Arctic Inference.

We would like to thank Flex Wang, Jerry Luo, Seth Li, Ricardo Aravena, Allen Woo and Vincent Chan for their continued support in bringing our research to production and into Snowflake Cortex AI.


\bibliographystyle{IEEEtran}
\bibliography{arctic_refs}

\begin{thebibliography}{10}
\providecommand{\url}[1]{#1}
\csname url@samestyle\endcsname
\providecommand{\newblock}{\relax}
\providecommand{\bibinfo}[2]{#2}
\providecommand{\BIBentrySTDinterwordspacing}{\spaceskip=0pt\relax}
\providecommand{\BIBentryALTinterwordstretchfactor}{4}
\providecommand{\BIBentryALTinterwordspacing}{\spaceskip=\fontdimen2\font plus
\BIBentryALTinterwordstretchfactor\fontdimen3\font minus \fontdimen4\font\relax}
\providecommand{\BIBforeignlanguage}[2]{{%
\expandafter\ifx\csname l@#1\endcsname\relax
\typeout{** WARNING: IEEEtran.bst: No hyphenation pattern has been}%
\typeout{** loaded for the language `#1'. Using the pattern for}%
\typeout{** the default language instead.}%
\else
\language=\csname l@#1\endcsname
\fi
#2}}
\providecommand{\BIBdecl}{\relax}
\BIBdecl

\bibitem{arcticInferenceRepo}
{Snowflake AI Research}, ``{ArcticInference: A vLLM plugin for low-latency, high‐throughput LLM inference},'' \url{https://github.com/snowflakedb/ArcticInference}, 2025.

\bibitem{vllmrepo}
{The vLLM Team}, ``{vLLM: A high-throughput and memory-efficient inference and serving engine for LLMs},'' \url{https://github.com/vllm-project/vllm}, 2025.

\bibitem{shiftParallelBlog2025}
S.~Rajbhandari, M.~Hidayetoglu, A.~Qiao, Y.~Wang, J.~Yang, J.~Rasley, and Y.~He, ``Arctic inference with shift parallelism: The fastest open source inference,'' \url{https://www.snowflake.com/en/engineering-blog/arctic-inference-shift-parallelism/}, May 2025.

\bibitem{specDecodeBlog2025}
{Snowflake AI Research}, ``Fastest speculative decoding in v{LLM} with {Arctic Inference and Arctic Training},'' \url{https://www.snowflake.com/en/engineering-blog/fast-speculative-decoding-vllm-arctic/}, June 2025.

\bibitem{swiftkvblog}
------, ``{SwiftKV}: Accelerating enterprise {LLM} workloads,'' \url{https://www.snowflake.com/en/engineering-blog/swiftkv-llm-compute-reduction/}, Dec. 2024.

\bibitem{embeddingBlog2025}
C.~Xu, J.~Yang, J.~Luo, D.~Campos, Y.~He, and S.~Rajbhandari, ``Scaling v{LLM} for embeddings: 16× throughput and cost reduction,'' \url{https://www.snowflake.com/en/engineering-blog/embedding-inference-arctic-16x-faster/}, June 2025.

\bibitem{teirepo}
{Hugging Face}, ``Text embeddings inference ({TEI}): A blazing fast inference solution for text embeddings models,'' \url{https://github.com/huggingface/text-embeddings-inference}, 2025.

\bibitem{ulyssesBlog2025}
M.~Hidayetoglu, A.~Qiao, J.~Rasley, Y.~He, and S.~Rajbhandari, ``Ulysses: Unlocking low-latency, high-throughput inference for long context {LLM}s,'' \url{https://www.snowflake.com/en/engineering-blog/ulysses-low-latency-llm-inference/}, Apr. 2025.

\bibitem{eagle2025}
\BIBentryALTinterwordspacing
Y.~Li, F.~Wei, C.~Zhang, and H.~Zhang, ``{EAGLE}: Speculative sampling requires rethinking feature uncertainty,'' 2025. [Online]. Available: \url{https://arxiv.org/abs/2401.15077}
\BIBentrySTDinterwordspacing

\bibitem{sglangrepo}
{The SGLang Team}, ``{SGL: a fast serving framework for large language models and vision language models},'' \url{https://github.com/sgl-project/sglang}, 2025.

\bibitem{vllmpaper}
\BIBentryALTinterwordspacing
W.~Kwon, Z.~Li, S.~Zhuang, Y.~Sheng, L.~Zheng, C.~H. Yu, J.~E. Gonzalez, H.~Zhang, and I.~Stoica, ``Efficient memory management for large language model serving with {PagedAttention},'' in \emph{Proceedings of the ACM SIGOPS 29th Symposium on Operating Systems Principles}, 2023. [Online]. Available: \url{https://arxiv.org/abs/2309.06180}
\BIBentrySTDinterwordspacing

\bibitem{sglangpaper}
\BIBentryALTinterwordspacing
L.~Zheng, L.~Yin, Z.~Xie, C.~Sun, J.~Huang, C.~H. Yu, S.~Cao, C.~Kozyrakis, I.~Stoica, J.~E. Gonzalez, C.~Barrett, and Y.~Sheng, ``{SGLang}: Efficient execution of structured language model programs,'' 2024. [Online]. Available: \url{https://arxiv.org/abs/2312.07104}
\BIBentrySTDinterwordspacing

\bibitem{trtllmrepo}
{NVIDIA Corporation}, ``{TensorRT‑LLM},'' \url{https://github.com/NVIDIA/TensorRT-LLM}, 2025.

\bibitem{swiftkvpaper}
\BIBentryALTinterwordspacing
A.~Qiao, Z.~Yao, S.~Rajbhandari, and Y.~He, ``{SwiftKV}: Fast prefill‐optimized inference with knowledge‐preserving model transformation,'' 2024. [Online]. Available: \url{https://arxiv.org/abs/2410.03960}
\BIBentrySTDinterwordspacing

\bibitem{sharegpt2023}
{anon8231489123}, ``{ShareGPT Vicuna Unfiltered: Cleaned English ShareGPT Conversations},'' \url{https://huggingface.co/datasets/anon8231489123/ShareGPT_Vicuna_unfiltered}, 2023.

\bibitem{humaneval2021}
\BIBentryALTinterwordspacing
M.~Chen, J.~Tworek, H.~Jun, Q.~Yuan, H.~P. de~Oliveira~Pinto, J.~Kaplan, H.~Edwards, Y.~Burda, N.~Joseph, G.~Brockman, A.~Ray, R.~Puri, G.~Krueger, M.~Petrov, H.~Khlaaf, G.~Sastry, P.~Mishkin, B.~Chan, S.~Gray, N.~Ryder, M.~Pavlov, A.~Power, L.~Kaiser, M.~Bavarian, C.~Winter, P.~Tillet, F.~P. Such, D.~Cummings, M.~Plappert, F.~Chantzis, E.~Barnes, A.~Herbert-Voss, W.~H. Guss, A.~Nichol, A.~Paino, N.~Tezak, J.~Tang, I.~Babuschkin, S.~Balaji, S.~Jain, W.~Saunders, C.~Hesse, A.~N. Carr, J.~Leike, J.~Achiam, V.~Misra, E.~Morikawa, A.~Radford, M.~Knight, M.~Brundage, M.~Murati, K.~Mayer, P.~Welinder, B.~McGrew, D.~Amodei, S.~McCandlish, I.~Sutskever, and W.~Zaremba, ``Evaluating large language models trained on code,'' 2021. [Online]. Available: \url{https://arxiv.org/abs/2107.03374}
\BIBentrySTDinterwordspacing

\bibitem{swebench2023}
\BIBentryALTinterwordspacing
C.~E. Jim{\'e}nez, J.~Yang, A.~Wettig, S.~Yao, K.~Pei, O.~Press, and K.~Narasimhan, ``{SWE}-bench: Can language models resolve real-world {GitHub} issues?'' 2023. [Online]. Available: \url{https://arxiv.org/abs/2310.06770}
\BIBentrySTDinterwordspacing

\bibitem{llama32024}
\BIBentryALTinterwordspacing
{Llama Team}, ``The llama 3 herd of models,'' \emph{arXiv preprint arXiv:2407.21783}, 2024. [Online]. Available: \url{https://arxiv.org/abs/2407.21783}
\BIBentrySTDinterwordspacing

\bibitem{arcticEmbed2024}
\BIBentryALTinterwordspacing
L.~Merrick, D.~Xu, G.~Nuti, and D.~Campos, ``Arctic‑embed: Scalable, efficient, and accurate text embedding models,'' 2024. [Online]. Available: \url{https://arxiv.org/abs/2405.05374}
\BIBentrySTDinterwordspacing

\bibitem{bgeBase2024}
{Beijing Academy of Artificial Intelligence}, ``{bge‑base‑en‑v1.5}: English text embedding model,'' \url{https://huggingface.co/BAAI/bge-base-en-v1.5}, 2024.

\end{thebibliography}

\section*{Appendix: Evaluation methodology}

\paragraph{Hardware:} All experiment results presented in this paper, unless otherwise stated, were run on an 8$\times$H200 GPU node, leveraging all the GPUs within the node.

\paragraph{Models:} Meta Llama 3.3 70B generative AI~\cite{llama32024}. Arctic Embedding model~\cite{arcticEmbed2024}, \href{https://huggingface.co/BAAI/bge-base-en-v1.5}{bge-base-en-v1.5}~\cite{bgeBase2024} for embedding.

Request completion latency and TPOT measurements: Unless otherwise stated, all request completion latency and TPOT measurements were done by sending one request at a time and averaged over a mixed data set consisting of ShareGPT, HumanEval and SWEBench, comprising short conversations, coding tasks and long agentic workflows. We used latency-optimized configs for all our baselines, where we used $\textrm{TP}=8$ and the best available open source speculative decoding approach supported by the baseline. 

\paragraph{Combined throughput measurements:} Unless otherwise stated, for throughput measurements all requests were sent at the same time and measurements were made ensuring steady state. For Figure 1, we constructed a realistic data set with input and output lengths sampled from our Snowflake Cortex AI production logs, allowing us to create a data set representative of enterprise workloads we see at Snowflake. For all other experiments, we used input length of 2,000 and output of 250 tokens, to match the average 10:1 ratio between input and output we observe in our production logs. We used throughput-optimized configs for all our baselines, where we used $\textrm{TP}=1$ and manually tuned the batch size to achieve the highest throughput.

\paragraph{Prefill throughput measurements:} Unless otherwise stated, prefill throughput measurements were performed using a single request with a 4,000 context length. This was because we found that context length smaller than 4,000 did not saturate the GPU, while a length longer than 4,000 did not increase prefill throughput significantly.  

\paragraph{Open source baselines:} vLLM (\texttt{v0.8.4}), TRT-LLM (\texttt{v0.18.2}), SGLang (\texttt{v0.4.6}) 

\paragraph{Open source speculative decoding baselines:} EAGLE-based speculative decoding offers the best latency for SGLang and vLLM, but it only supports shorter than 4,000 sequences and crashes when longer sequences are sent. Hence we could not use it for our realistic data set mix, which consisted of both short and longer sequences. Due to the limitations of EAGLE, we leveraged NGRAM speculation in vLLM and no speculation in SGLang, as anything else could not support the real-world use case we see in our production. For TRT-LLM, despite our best effort, we could not get it to work with any speculative decoding in a consistent way, and therefore we reported the numbers without speculative decoding.

\paragraph{Arctic Inference+vLLM:} Unless otherwise stated, for all experiments referred to as Arctic Inference, we used a single config combining Shift Parallelism shifting between $\textrm{SP}=8$ (Arctic Ulysses) and $\textrm{TP}=8$, with \href{https://www.snowflake.com/en/engineering-blog/swiftkv-llm-compute-reduction/}{SwiftKV} optimizations, and \href{https://www.snowflake.com/en/engineering-blog/fast-speculative-decoding-vllm-arctic/}{speculative decoding optimizations} that combine suffix decoding with our LSTM draft model. We ran Arctic Inference on top of vLLM (\texttt{v0.8.4}).

\end{document}